\documentclass[aps,showpacs,amsmath,amssymbd,dvips,showkeys,preprint]{revtex4}

\usepackage{graphicx}
\usepackage{graphics}
\usepackage{color}
\def \be  {\begin{equation}}
\def \ee  {\end{equation}}
\def \ee  {\end{equation}}
\def \bea {\begin{eqnarray}}
\def \eea {\end{eqnarray}}

\begin{document}

\preprint{ECTP-2013-13\hspace*{0.5cm}and\hspace*{0.5cm}WLCAPP-2013-10}

\title{Koppe's Work of 1948\footnote{H. Koppe, {\it Die Mesonenausbeute beim Beschuss von leichten Kernen mit Alpha-Teilchen}, Z. Naturforsch. {\bf 3a}, 231-232 (1948); \newline
[English Translation] Meson Yields from Bombarding Light Nuclei with Alpha Particles, Z. Naturforsch. A {\bf xx}, xxx (2013)}:\\
A fundamental for non-equilibrium rate of particle production}
\author{Abdel Nasser TAWFIK\footnote{http://www.atawfik.net/, ~email: a.tawfik@eng.mti.edu.eg}}
\affiliation{Egyptian Center for Theoretical Physics (ECTP), MTI University, 11571 Cairo, Egypt}
\affiliation{World Laboratory for Cosmology And Particle Physics (WLCAPP), Cairo, Egypt}

\date{\today}

\begin{abstract}
In 1948, Koppe formulated an almost complete recipe for statistical-thermal models including particle production, formation and decay of resonances, temporal and thermal evolution of the interacting system, statistical approaches and equilibrium condition in final state of the nuclear interaction. As the rate of particle production was one of the basic assumptions, recalling Koppe's work would be an essential input to be involved in the statistical prediction of non-equilibrium particle production in recent and future ultra-relativistic collisions. 
\end{abstract}


\maketitle

\section{Rate of particle production in thermal medium}

At Betatron energy, the temperature of excited nucleus $T_0=3.8\sqrt{N}$, where $N$ refers to the number of excited nuclei (resonances), was related to the excitation energy $U=m_2/(m_1+m_2)^2\, E$, with $m_1 (m_2)$ and $E$ being mass of projectile (mass of target) nucleus and kinetic energy, respectively. $T_0$ was measured as $\sim 10~$MeV. Koppe assumed that even the projectile ($\alpha$-particle) can not remain stable \cite{hkoppe}. Therefore, pair production is likely \cite{hkoppe2}. {\it ''pair-degeneracy''} \cite{hkoppe2} or {\it ''Vacuum dissociation''} \cite{refffff2} was principally investigated from electron-positron pair production. Koppe assumed that the same considerations make it possible to apply this in meson-pair production \cite{hkoppe} and expected a very small number of produced mesons. 

The electron (produced particle) gas can be treated as cavity radiation with a concrete energy density relating  radiation loss to cross-section of excited nuclei $\sigma$. The temporal evolution of temperature should reflect the expansion of the interacting system. The speed of electrons (produced particles) is very close to $c$. Then, the energy flux caused by electrons relative to light quantum radiation will be increased by the same factor $7/8$. The rate of produced particle was calculated as, 
\bea
\nu(T(t)) &=& \frac{m_b\, \sigma}{\pi^2\, \hbar^3}\, T(t)^2\, e^{-\frac{m_b\, c^2}{T(t)}}.
\eea
Then, the integration results in $ n=a (m_1+m_2) \, T_0 \, \exp(-m_b\, c^2/T_0)$,
where $a=0.031$. Substituting with given values of energy and $T_0$, then the number of mesons which should be produced in $\alpha-A$ collisions at $380~$MeV per unit time was estimated as $\sim1.7\times10^{-4}$. 

\section{Particle production and non-equilibrium particle distribution}

In ultra-relativistic nuclear collisions, deconfinement and/or chiral broken-symmetry restoration phase transition(s) is(are) supposed to take place. To study the dynamics and velocity distribution of objects in such thermal background, like transport properties in quark-gluon plasma, Fokker-Planck equation is a well-known tool. The statistical properties of an ensemble consisting of individual parton objects is given by non-equilibrium single-particle distribution function $f$ \cite{Tawfik:2010kz,Tawfik:2010uh,Tawfik:2010pt}. The probability of finding an object in infinitesimal region in phase space is directly proportional to the volume element and $f$. The latter is assumed to fulfil the Boltzmann-Vlasov (BV) master equation,
\bea \label{eq:te1}
\dot f + \dot{\vec{x}} \cdot \nabla_x f + \dot{\vec{k}} \cdot \nabla_k f + \dot{\vec{q_c}} \cdot \nabla_{q_c} f &=& {\cal G}  + {\cal L}.
\eea
The first term in rhs ${\cal G}$ represents gain or rate of particle production with momentum $k + kt$, which is conjectured to lose momentum $kt$ due to reactions with the background. The second term ${\cal L}$ represents loss due to the scattering rate. The effective potential $U$ has to combine the well-known Coulomb $U(x\rightarrow 0)\propto 1/x$ and confined potentials $U(x\rightarrow \infty)\propto 0$. The standard position $\vec{x}$ and momentum $\vec{k}$ variables are given in first two terms in lhs. The third term represents the dynamics of the charge, where $\dot{\vec{k}}$ can be given by field tensor. The fourth term reflects an extension of phase space to include color charge, $\dot{\vec{q_c}}$. 

Studying the stochastic behavior of a single object propagating with random noise known as Langevin equation represents one way to solve this problem. A master equation, such as the linearised  BV equation, with the Landau soft-scattering approximation would give another method. Koppe's work would be a fundamental-statistical approach for a qualitative estimation for the non-equilibrium rate of particle production.


\begin{thebibliography}{00}  %for 3 digits

\bibitem{hkoppe} H. Koppe, Z. Naturforschg. {\bf 3 a}, 251-252 (1948).

\bibitem{hkoppe2} H. Koppe, Ann. Phys.(Berlin) {\bf 437}, 103-112 (1948).

\bibitem{refffff2} F. Houtermans and H. Jensen, Z. Naturforsch. {\bf 2 a}, 146  (1947).


\bibitem{Tawfik:2010kz} A. Tawfik, %{\it ''Matter-antimatter asymmetry in heavy-ion collisions''},  
Int. J. Theor. Phys. {\bf 51}, 1396-1407 (2012).%;e-Print: arXiv:1011.6622 [hep-ph].

\bibitem{Tawfik:2010uh} A. Tawfik, %{\it ''Dynamical Fluctuations in Baryon-Meson Ratios''}, 
J.Phys. {\bf G40}, 055109 (2013).%; arXiv:1007.4585 [hep-ph].

\bibitem{Tawfik:2010pt} A. Tawfik, %{\it ''Antiproton-to-Proton Ratios for ALICE Heavy-Ion Collisions''}, 
Nucl. Phys. A {\bf 859}, 63-72 (2011).%, e-Print: arXiv:1011.5612 [hep-ph].

\end{thebibliography}
\end{document}